\documentclass[aps, superscriptaddress, amsmath,amssymb,reprint]{revtex4-2}
\usepackage{graphicx}
\usepackage{dcolumn}
\usepackage{bm}
\usepackage[utf8]{inputenc}
\usepackage[T1]{fontenc}
\usepackage{mathptmx}
\usepackage{etoolbox}
\usepackage{xcolor}
\usepackage{url}
\usepackage{natbib}
\usepackage{hyperref}
\usepackage{subcaption}
\makeatletter
\def\@email#1#2{%
 \endgroup
 \patchcmd{\titleblock@produce}
  {\frontmatter@RRAPformat}
  {\frontmatter@RRAPformat{\produce@RRAP{*#1\href{mailto:#2}{#2}}}\frontmatter@RRAPformat}
  {}{}
}%
\makeatother
\begin{document}


\title{Revisiting multifunctionality in reservoir computing}

\author{Swarnendu Mandal}
\email{swarnenduphy35@gmail.com}
\affiliation{International Research Center for Neurointelligence (WPI-IRCN), The University of Tokyo, Tokyo, Japan}

\author{Kazuyuki Aihara}
\affiliation{International Research Center for Neurointelligence (WPI-IRCN), The University of Tokyo, Tokyo, Japan}

\date{\today}

\newcommand{\blue}{\color{blue}}
\newcommand{\red}{\color{red}}

\begin{abstract}

Multifunctionality is ubiquitous in biological neurons. Several studies have translated the concept to artificial neural networks as well. Recently, multifunctionality in reservoir computing (RC) has gained the widespread attention of researchers. Multistable dynamics of the reservoir can be configured to capture multiple tasks, each by one of the co-existing attractors. However, there are several limitations in the applicability of this approach. So far, multifunctional RC has been shown to be able to reconstruct different attractor climates only when the attractors are well separated in the phase space. We propose a more flexible reservoir computing scheme capable of multifunctioning beyond the earlier limitations. The proposed architecture holds striking similarity with the multifunctional biological neural networks and showcases superior performance. It is capable of learning multiple chaotic attractors with overlapping phase space. We successfully train the RC to achieve multifunctionality with wide range of tasks.
\end{abstract}

\maketitle


\section{\label{sec:introduction}Introduction}
The concept of multifunctionality has been studied in neuroscience since the 1980s, with early foundational work by Mpitsos and Cohan \cite{mpitsos1986convergence}, and later reviews by Dickinson \cite{dickinson1995interactions}, Marder and Calabrese \cite{marder1996principles}. More recent studies by Briggman and Kristan have expanded our understanding of how multifunctionality emerges in biological systems \cite{briggman2008multifunctional}. Multifunctionality in neural networks refers to the ability of a neural circuit to switch between different dynamical behaviors or tasks without altering its synaptic properties. This means that the same network of neurons can generate different behaviors based on external inputs or conditions rather than needing to modify the connections (synapses) between neurons. This property is crucial in biological neural networks (BNNs), particularly in systems where a small number of neurons control multiple, mutually exclusive behaviors \cite{getting1989emerging}. One well-studied example shows the ability of certain organisms to switch between swimming and crawling using the same neural circuit \cite{briggman2006imaging,kristan1988multifunctional}. Moreover, in humans, the regulation of distinct breathing patterns such as regular breathing, sighing, and gasping within the same respiratory network is another instance of multifunctionality by real neural networks \cite{lieske2000reconfiguration}. 

Multifunctionality is not limited to biological systems; it has important implications for artificial neural networks (ANNs) and machine learning (ML) as well \cite{yang2019study}. In traditional ANNs, different tasks often require separate networks or retraining of weights. Multifunctionality suggests that a single ANN could perform multiple tasks using the same trained weights by leveraging multistability. This can unlock additional computational capabilities, allowing a network to store and recall multiple functions more efficiently. By integrating multistability principles into ML, researchers can design networks that dynamically switch between different operational modes without retraining \cite{yang2019task,hong2019training,benson2020neural}. This could significantly improve the flexibility and adaptability of AI models.

One key application of multifunctionality in machine learning is in Reservoir Computing (RC), a computational framework inspired by dynamical systems \cite{jaeger2002tutorial,lukovsevivcius2009reservoir,tanaka2019recent}. RC leverages a high-dimensional dynamical system as a reservoir to process inputs without requiring extensive training \cite{yan2024emerging}. Numerous studies exploited the capabilities of the RC framework to advance the understanding of dynamical system \cite{grigoryeva2018echo,kong2023reservoir,pathak2018model,lu2017reservoir,pathak2017using,lu2018attractor, vlachas2020backpropagation,zhang2021learning,fan2021anticipating,xiao2021predicting,roy2022model,mandal2023learning}. Multistability is an abundant phenomenon in dynamical systems which may contain several coexisting attractors \cite{pisarchik2014control,feudel1997multistability}. Depending on initial conditions, the dynamical system can evolve into one of these attractors. By training a reservoir network to reconstruct these co-existing attractors, one can create multifunctional RCs capable of handling multiple tasks within a single model. This was the original idea behind multifunctionality in RC, proposed by Flynn et al \cite{flynn2021multifunctionality}. However, there are several limitations to the applicability of the approach. One of the major issues is that the machine struggles to multifunction with the tasks involving attractors sharing common phase space \cite{flynn2023seeing}. Other challenges include narrow multifunctionality regions in the network's parameter space and restrictions on the time scale difference of the learned attractors, etc. A recent study \cite{du2025multifunctional} attempted to broaden the range of applications of multifunctional RC by inserting an additional scalar `label' input to distinguish attractors. However, this method too relies upon a manual separation of trajectories of the attractors in the phase space.

\begin{figure}
\includegraphics[width = \linewidth]{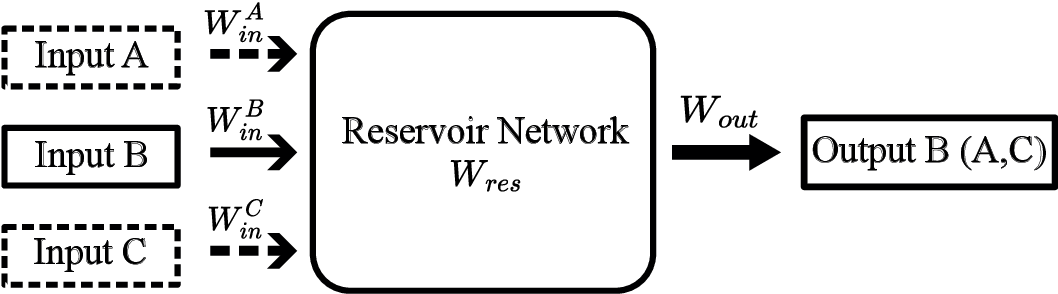}
\caption{\label{fig:scheme}Schematic of the proposed multifunctional reservoir computing architecture.
}
\end{figure}

In this work, we propose an architecture for multifunctional RC that presents superior functionality and broader applicability. In this architecture, separated input channels are employed to take input for different tasks as shown in Figure \ref{fig:scheme}. These inputs channels act as controls for selecting a particular task for the reservoir to perform. While performing one of the tasks, only the corresponding input channel participates, and others remain inactive. Multiple studies on multifunctional BNNs suggest that there are separate control mechanisms to trigger different functions. In our scheme, the reservoir layer and the readout layer can be considered to contain neurons that multifunction and participate in all the tasks. Only the input layer contains task-specific neurons. Thus, the proposed scheme seems biologically more relevant as well. This extension of RCs enhances their computational power, enabling them to model complex behaviors efficiently and adapt to new tasks without modifying core parameters.

In the following section (Sec. \ref{sec:background}), we highlight the limitations of the earlier approach for multifunctional RC and justify the proposed architecture from the examples of real-life multifunctional BNNs. Sec. \ref{sec:arch} describes the architecture in details. In Sec. \ref{sec:numsim}, we provide results of numerical experiments for multiple scenarios of the reservoir multifunctioning. For all the cases, we have considered the tasks involving attractors with overlapping phase space. Finally, in Sec. \ref{sec:disc}, we give a critical review of the architecture after summarizing the results and then conclude.

\section{Background}\label{sec:background}

\subsection{\label{ssec:limit_flynn}Challenges in previous approach}

Here, we highlight several limitations regarding the applicability of the previously proposed scheme of multifunctionality in RC.

\paragraph{Challenges in Distinguishing Overlapping Attractors:} When multiple attractors share common phase space, the RC struggles to maintain distinct representations, requiring higher memory capacity to separate them. The ``seeing double" problem illustrates how overlapping training data forces the RC to use more memory, making attractor reconstruction increasingly difficult \cite{flynn2023seeing}. This can lead to errors in classification and affect the system’s robustness in performing different tasks. The system may favor reconstructing one attractor over the other, leading to decreased effectiveness.

%

\paragraph{Dependency on Timescales:} The RC’s ability to reconstruct multiple attractors is sensitive to their relative timescales. The reservoir's internal dynamics struggles to accommodate both fast and slow attractors simultaneously, limiting its ability to switch between tasks smoothly. If one attractor evolves on a significantly different timescale than another, multifunctionality may be lost altogether.

\paragraph{Sensitivity to Network Parameters:}
The success of multifunctionality depends heavily on parameters like spectral radius ($\rho$) and network size ($N_{res}$) of the reservoir. Only a narrow range of $\rho$ values enables proper attractor reconstruction. Changes in $\rho$, connectivity sparsity, and input strength can lead to bifurcations that destabilize performance, limiting functional reliability.

\paragraph{Untrained Attractors in Prediction Space:} Even when trained for multifunctionality, additional ``untrained attractors” can emerge within the prediction state space. These attractors are not explicitly part of the training but can interfere with the desired attractor reconstruction, reducing the reliability of the system.

These limitations indicate that while the approach by Flynn et al. is a promising one to achieve multifunctionality in reservoir computing, its effectiveness depends on precise tuning of system parameters and a careful selection of attractors that do not overly interfere with one another.

\subsection{Motivation from mutifunctional BNNs}\label{ssec:bnn}

In this section, we justify our approach with examples from different multifunctional biological neural networks. We discuss how separate input channels are employed to control different multifunctional behaviors. This motivates our proposed modifications in RC architecture to achieve superior multifunctional performance with broader applicability. 

\paragraph*{Presence of Task-specific Neurons:}
The findings from Briggman, Kristan, and William (2006) \cite{briggman2006imaging} suggest that while some neurons in the central pattern generators (CPGs) of the medicinal leech are multifunctional, others are dedicated to specific behaviors. The study identified both multifunctional and dedicated neurons in the networks controlling swimming and crawling. Cell 255 was shown to be multifunctional, participating in both swimming and crawling and influencing both behaviors. Cell 257, in contrast, was found to be a dedicated interneuron, participating only in crawling and becoming hyperpolarized (inactive) during swimming. The existence of dedicated neurons supports the idea that separate channels control different behaviors.

\paragraph*{Distinct Neural Pathways for Different Functions:}
In Tritonia diomedea, Dorsal Swim Interneurons (DSIs) excite both the swimming and crawling circuits, acting as an example of multifunctional BNN \cite{popescu2002highly}. But they trigger two actions in different ways. Swimming is controlled by a rhythmic CPG, while crawling is regulated by cilia-activating neurons (Pd21 and Pd5), suggesting functionally distinct pathways for controlling multifunctionality.

\paragraph*{Selective Neuromodulation:}
The pre-Bötzinger complex shows that breathing behaviors are modulated by different control mechanisms \cite{lieske2000reconfiguration}. The ability of the same neural network to generate different breathing patterns is due to changes in the balance of excitatory and inhibitory synaptic inputs.
Eupnea is characterized by a balance between excitatory and inhibitory inputs, leading to a stable rhythmic pattern. Sighs involve an enhancement of excitatory inputs, leading to a biphasic burst. Gasping is associated with a reduction in synaptic inhibition, allowing excitatory inputs to dominate and producing a more abrupt and shorter burst.

\section{\label{sec:arch}The Architecture}

Consider a multifunctional echo-state network capable of handling $N_{task}$ tasks. The reservoir network is defined by a sparse adjacency matrix $W_{res}$. Each node of the network possesses a state typically initiated at zero. The state of the whole network at any instant $t$ is defined by a column matrix $\bm{x}_t$, where each element of this matrix represents the state of the individual node. The reservoir dynamics, driven by an input signal $\bm{u}^i(t)$ is given in Equation (\ref{eq:esn}). The input is fed to the reservoir by an input connection matrix $W^i_{in}$.  $\bm{u}^i_t$ is the input vector at instant $t$.

\begin{equation}\label{eq:esn}
\bm{x}_{t+1} = (1 - \alpha)\bm{x}_t + \alpha {\rm tanh}(W^i_{in} \bm{u}^i_t + W_{res} \bm{x}_t).
\end{equation}
Here, $\alpha$ is called the leaking parameter, a hyperparameter for the setup. The shape of the matrices used is described in Table \ref{tab:dim}. The superscript index $i$ represents different tasks, $i = 1,~2,~\ldots,~N_{task}$.

In the proposed scheme, the input connection matrices are prepared in a special way. Each of the tasks has its own dedicated input connection matrix. The reservoir nodes are divided into $N_{task}$ groups, and each group is assigned to take input for only one task. The total number of reservoir nodes $N_{res}$ can be expressed as $N_{res} = \sum_{i = 1}^{N_{task}} N^i$, where $N^i$ is the number of nodes responsible for accepting input for the $i^{\rm th}$ task. This way, any of the input matrices $W^i_{in}$ does have only $N^i$ randomly assigned nonzero rows corresponding to its dedicated reservoir nodes, which are mutually exclusive to other input matrices. In general, the values of $N^i$ ($i = 1,~2,~\ldots,~N_{task}$) may not be equal to each other. The values can be decided depending on the requirements considering the relative complexity of the corresponding task. Moreover, if the random nonzero element of $W^i_{in}$ is drawn from the interval $[-\sigma^i,\sigma^i]$, the value of $\sigma^i$ can also be optimized depending upon the memory requirement of the particular task. An important point to note here is that a group of nodes is specific to a task in terms of input-receiving nodes only. However, since the connections are recurrent, all of the reservoir nodes participate in reservoir functioning for any task.

 The reservoir dynamics for the $i^{\rm th}$ task is stored in $\mathfrak{R}^i$  after squaring the alternate elements in each reservoir state.
 
 \begin{equation}\label{eq:asym}
 [\mathfrak{r}_t]_j = 
 	\begin{cases}
        [x_t]_j & \text{if }j \text{ is odd},\\
        [x_t^2]_j & \text{if } j \text{ is even},
    \end{cases} 
 \end{equation}
where, $\mathfrak{r}_t$ represents one column of $\mathfrak{R}^i$ corresponding to instant $t$. $\mathfrak{r}_t$ can be mapped to the target output $\bm{v}_t$ as

\begin{equation}\label{eq:readout}
\bm{v}_t = W_{out} \mathfrak{r}_t,
\end{equation}
where $W_{out}$ is the output connection matrix, the only trainable part of this setup. $W_{out}$ is evaluated in training phase using Equation (\ref{eq:regression}). Specifically for the task of attractor climate reconstruction, $\bm{v}_t$ is used as the ESN input for the next step $\bm{v}_t = \bm{u}^i_{t+1}$ in Equation (\ref{eq:esn}). Thus, the reservoir works in a closed-loop setup, producing the dynamics of the attractor with an arbitrary number of time steps.

\begin{table}[h!]
\begin{tabular}{|c|c|c|}
\hline
 Matrix & Dimension & Description \\ 
\hline 
\hline
 $W_{res}$ & $N_{res} \times N_{res}$ & $N_{res}$ - number of reservoir neurons. \\ 
\hline 
$W^i_{in}$ & $N_{res} \times D^i_u$ & $D^i_u$ - dimension of input for the $i^{\rm th}$ task\\ 
\hline 
$\mathfrak{R}^i$ & $N_{res} \times T^i_{tr}$ & $T^i_{tr}$ - training data length for the $i^{\rm th}$ task \\ 
\hline 
$\mathfrak{R}$ & $N_{res} \times T_{tr}$ & $T_{tr} = \sum_{i = 0}^{N_{task}} T^i_{tr} $ \\ 
\hline 
$V^i$ & $D^i_v \times T^i_{tr}$ & $D^i_v$ - dimension of output for the $i^{\rm th}$ task\\ 
\hline 
$V$ & $D^{max}_v \times T_{tr}$ & $D^{max}_v$ - highest value of output dimensions\\ 
\hline 
$\mathbb{I}$ & $N_{res} \times N_{res}$ & $\mathbb{I}$ - identity matrix\\ 
\hline
$W_{out}$ & $D^{max}_v \times N_{res}$ & $W_{out}$ - trainable output weight\\ 
\hline 
\end{tabular} 
\caption{Shape and size of the matrices used in the echo-state network.}
\label{tab:dim}
\end{table}

During the training phase, the only objective is to evaluate $W_{out}$. For our multifunctional reservoir computing setup, we use the training data blending technique to train the reservoir \cite{flynn2021multifunctionality}. First, the reservoir state matrices $\mathfrak{R}^i \forall (i = 1,~2,~\ldots,~N_{task})$ are generated using training input data for all the tasks with the help of Equations (\ref{eq:esn}) and (\ref{eq:asym}). Then, a final reservoir state matrix is prepared by staking all of them together, as $\mathfrak{R} = [\mathfrak{R}^1~\mathfrak{R}^2~\cdots~\mathfrak{R}^{N_{task}}]$. The corresponding teacher matrix is prepared in the same way, using the training output data (labels). The training label corresponding to any training input data point is the data at the subsequent time step. Thus, the machine learns to produce the state variables at the next time steps from the input of the current state. However, preparing the final teacher matrix from the ones for individual tasks faces the dimension mismatch issue in blending. In general, individual tasks may have different input and output dimensions from the others. We consider the largest value of dimensions of all the tasks. In the teacher matrices corresponding to the task with a lower output dimension, we use additional zero rows to match it to the value of the largest. For example, if the largest output dimension is $D_v^{max}$, any task with output dimension $D_v^{max} - l$ will have an additional $l$ zero rows in its teacher matrix. Thus, we can prepare the final teacher matrix as $V = [V^1~V^2~\cdots~V^{N_{task}}]$. The output connection matrix is evaluated by Ridge regression given by

\begin{equation}\label{eq:regression}
W_{out} = (V\cdot \mathfrak{R}^T)\cdot (\mathfrak{R} \cdot \mathfrak{R}^T + \beta \mathbb{I})^{-1}.
\end{equation}

$\beta$ is called the regularization parameter, another hyperparameter of the setup. $\mathbb{I}$ is the identity matrix. 

The teacher matrix $V$ is a $D_v^{max} \times T_{tr}$ matrix where $T_{tr}$ is the sum of all training lengths for individual tasks, i.e. $T_{tr} = \sum_{i=1}^{N_{task}} T^i_{tr}$. The training length $T^i_{tr}$ is considered as the number of reservoir run steps with training data, after discarding the reservoir's transient dynamics. For all the tasks, the reservoir was initialized at zero, and enough reservoir transient was removed for all the tasks. The training data length plays an important role in the performance of any ESN. In our case, the training length for individual tasks, relative to the total training length, can also be optimized considering the complexity and requirements of individual tasks.

\section{Numerical simulations}\label{sec:numsim}

\subsection{Climate Reconstruction for Lorenz and R\"{o}ssler Systems}\label{ssec:numsim1}
We consider two characteristically distinct chaotic attractors to showcase the performance of our proposed method. The reservoir learns to multifunction to reconstruct the climate of Lorenz and R\"{o}ssler attractors. The Lorenz system is described by \cite{lorenz2017deterministic}

\begin{align}\label{eq:lorenz}
\dot{x_L} &= 10(y_L - x_L),  \nonumber \\
\dot{y_L} &=  - y_L + x_L(28 - z_L) ,  \\
\dot{z_L} &= x_Ly_L - \frac{8}{3}z_L,  \nonumber  
\end{align}

and the R\"{o}ssler system is given by \cite{rossler1976equation}

\begin{align}\label{eq:rossler}
\dot{x_R} &= - y_R - z_R,  \nonumber \\
\dot{y_R} &= x_R + 0.2y_R ,  \\
\dot{z_R} &= 0.2 + x_R(z_R-5.7).  \nonumber  
\end{align}

\begin{figure}
\includegraphics[width = \linewidth]{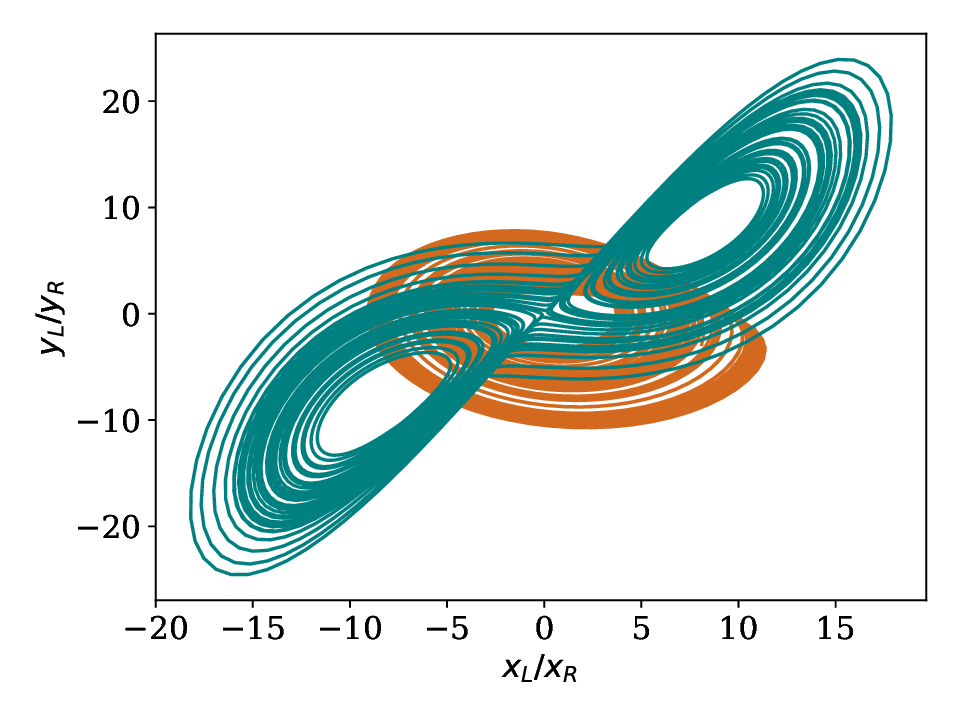}
\caption{\label{fig:ros-lor}Plot of multifunctionally reconstructed R\"{o}ssler and Lorenz attarctors.}
\end{figure}

These two systems have a considerable difference in their time scale. Moreover, the R\"{o}ssler attractor has a significantly higher memory requirement than that of Lorenz \cite{carroll2022optimizing}. Despite the numerical overlap in the phase space, our scheme successfully predicts both attractors' climates multifunctionally, as shown in Figure \ref{fig:ros-lor}.

There are three hyperparameters that control the quality of training of each attractor. These task specific hyperparameters are $N^i$, $\sigma^i$ and $T^i_{tr}$, where superscript $i$ denotes a task. Though Lorenz and R\"{o}ssler attractors have many qualitative differences, we have kept all these parameters identical for both tasks in this case. $N^1 = N^2 = N_{res}/2$,  $\sigma^1 = \sigma^2 = 0.1$ and $T^1_{tr} = T^2_{tr} = 50000$ when training data are generated from Equation (\ref{eq:lorenz} \& \ref{eq:rossler}) and sampled at $\Delta t = 0.01$ and $\Delta t = 0.1$ respectively. The rest of the hyperparameter values for the whole setup are given in Table \ref{tab:hype}.

\subsection{Climate of Systems with Different Dimensions}

We highlight the flexibility of our setup by training the reservoir to learn attractors with different dimensions for multifunctioning. Along with the Lorenz system described in Equation (\ref{eq:lorenz}), we consider a 5D hyper-chaotic Sprott B system given by \cite{ojoniyi20165d}

\begin{align}\label{eq:sprott5}
\dot{x_S} &= x_Sz_S - v_S,  \nonumber \\
\dot{y_S} &= x_S - y_S - u_S ,  \nonumber  \\
\dot{z_S} &= 1 - x_Sy_S, \\
\dot{u_S} &= 0.01x_S + y_S, \nonumber  \\
\dot{v_S} &= x_S. \nonumber
\end{align}

\begin{figure}
\includegraphics[width = \linewidth]{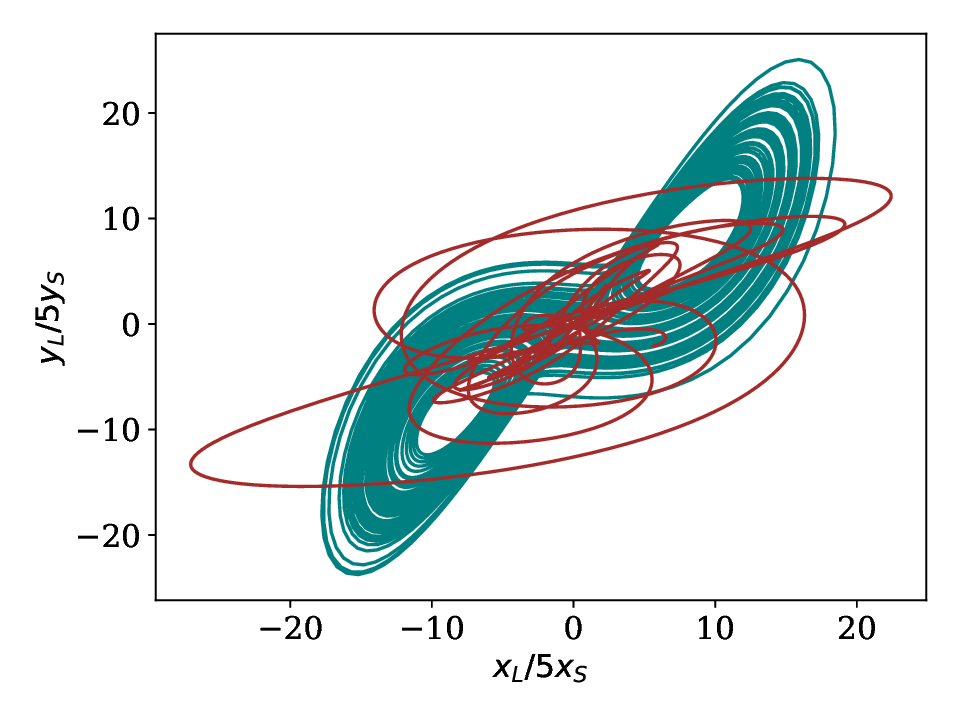}
\caption{\label{fig:lor-sp5}Plot of multifunctionally reconstructed Lorenz attractor and hyper-chaotic Sprott B attractor.}
\end{figure}

This system has two positive maximum Lyapunov exponents $\lambda_1 = 0.0981$ and $\lambda_2 = 0.0156$, making it hyper-chaotic. Figure \ref{fig:lor-sp5} shows two RC reconstructed attractors in a numerically common phase space. For better visualization in comparison with the Lorenz attractor, we have plotted the 5D Sprott B attractor in an enlarged space ($5x_S-5y_S$ plane).

To tackle the qualitative differences in the individual attractors, this time, we introduced asymmetry in the task-specific hyper-parameters. A higher number of reservoir nodes are assigned to take input for the system with higher dimensions. We distributed the reservoir nodes as $N^1:N^2 = 2:1$ for 5D Sprott B and Lorenz attractors, respectively. A larger training data set is also used in the case of the hyper-chaotic system. $T^1_{tr} = 70000$ and $ T^2_{tr} = 30000$ while both of the training data are generated from Equation (\ref{eq:lorenz} \& \ref{eq:sprott5}) sampling at $\Delta t = 0.01$. $\sigma^i$ is kept equal for both cases with $\sigma^1 = \sigma^2 = 0.1$.

\subsection{Climate Reconstruction of Two Hyperchaotic Systems}

The superiority in performance of our proposed multifunctional RC scheme is also evident in the case of reconstructing the climate of two attractors, both being hyper-chaotic. We consider the 5D Sprott B system described in Equation (\ref{eq:sprott5}) along with hyper-chaotic Laarem system described as \cite{laarem2021new}

\begin{equation}\label{eq:laarem}
\begin{aligned}
\dot{x_E} &= -y_E - z_E + w_E,  \\
\dot{y_E} &= x_E + 0.1y_E - z_E,  \\
\dot{z_E} &= 0.1 + z_E(x_E - 14) - 20x_Ew_E, \\
\dot{w_E} &= x_Ey_E -0.28w_E. \\
\end{aligned}
\end{equation}

\begin{figure}
\includegraphics[width = \linewidth]{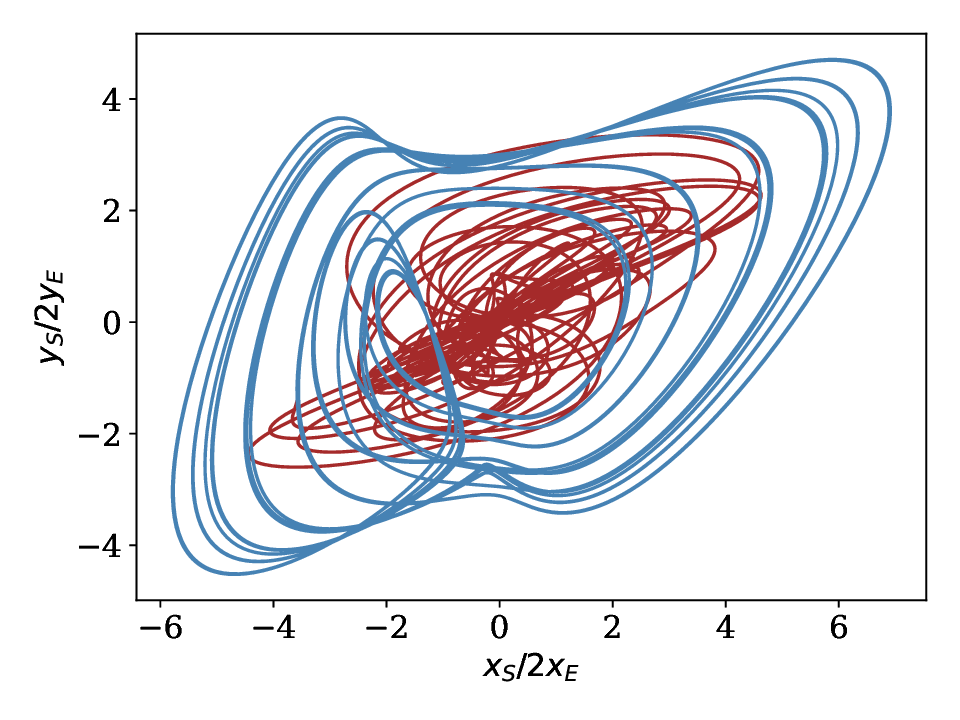}
\caption{\label{fig:sp5-laar}Plot of multifunctionally reconstructed hyper-chaotic 5D Sprott B attractor and hyper-chaotic Laarem attractor.}
\end{figure}

This attractor is hyper-chaotic for the chosen set of parameter values with two positive maximum Lyapunov exponents being $\lambda_1 = 0.5516$ and $\lambda_2 = 0.4059$. Figure \ref{fig:sp5-laar} depicts the performance of multifunctionality for this case. All the task-specific hyper-parameters are kept equal for both tasks as $N^1 = N^2 = N_{res}/2$,  $\sigma^1 = \sigma^2 = 0.1$ and $T^1_{tr} = T^2_{tr} = 50000$. Training data are generated from Equation (\ref{eq:sprott5} \& \ref{eq:laarem}) sampling with $\Delta t = 0.01$ for both cases.

\subsection{Three Climate Reconstruction Tasks}

The scalability of the scheme is judged by subjecting it to multifunction for three different climate reconstruction tasks. In addition to Lorenz and R\"{o}ssler system, Chua's attractor is introduced. Dynamics of the Chua system is given by \cite{chua1994chua}

\begin{equation}\label{eq:chua}
\begin{aligned}
\dot{x} &= \alpha(y-x-\phi(x)),\\
   \dot{y} &= x-y+z,\\
   \dot{z} &= -\beta y, \\
\end{aligned}
\end{equation}
 where, $\phi(x) = m_1x + \dfrac{1}{2}(m_0 - m_1)(|x+1|-|x-1|)$.

\begin{figure}
\includegraphics[width = \linewidth]{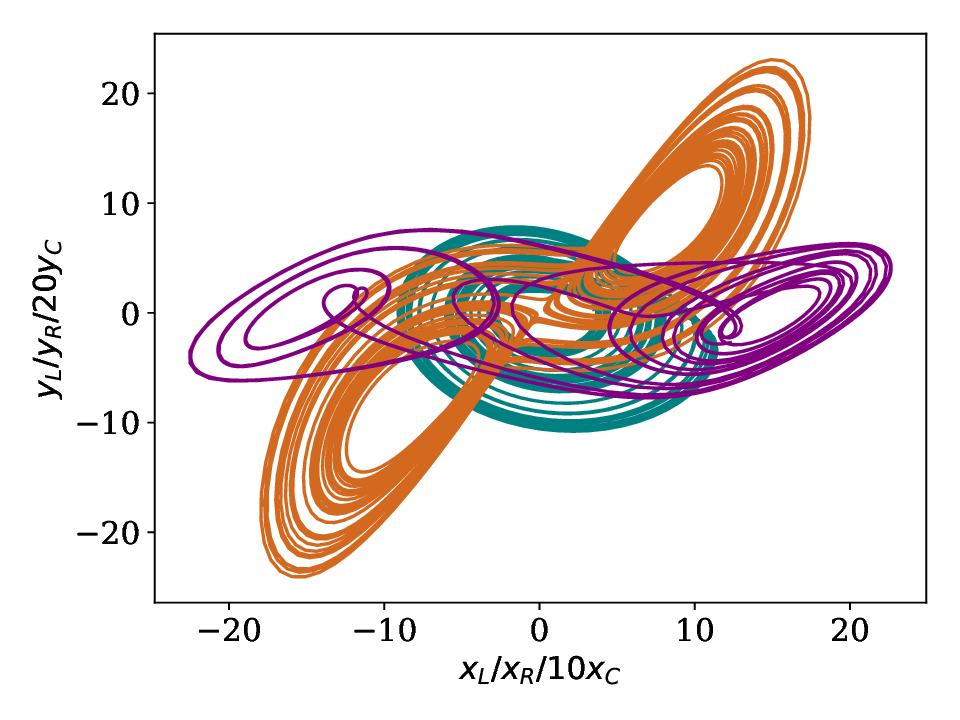}
\caption{\label{fig:lrc3}Plot of multifunctionally reconstructed Lorenz, R\"{o}ssler and Chua attractor. For better visualization, the Chua attractor is scaled ten times on the $x$-axis and twenty times on the $y$-axis.}
\end{figure}

The parameters are fixed at $\alpha = 15.6$, $m_0 = -8/7$, $m_1 = -5/7$ and $\beta = 28$, in order to have the chaotic attractor. The scheme is equally successful in showing multifunctionality with three tasks, as shown in Figure \ref{fig:lrc3}. $N^1 = N^2 = N^3 = N_{res}/3$ and $\sigma^1 = \sigma^2 = \sigma^3 = 0.1$ are the values of the hyperparameters. The training data lengths are set as $T^1_{tr} = 30000$, $T^2_{tr} = 50000$ and and $T^2_{tr} = 20000$ for Lorenz, R\"{o}ssler and Chua systems respectively. The training data set is prepared from Equations (\ref{eq:lorenz}, \ref{eq:rossler} and \ref{eq:chua}) with respective sampling rates $\Delta t = 0.01$, $\Delta t = 0.1$ and $\Delta t = 0.01$.

\subsection{Reservoir Observer for Lorenz and R\"{o}ssler Systems}\label{ssec:numsim5}

To showcase the versatility of multifunctioning, we consider another class of tasks, namely {\em reservoir observer} \cite{lu2017reservoir}. Unlike the climate reconstruction tasks, in this problem, a machine has to infer the missing variables of the system from the information of available state variables. In this case, we use Lorenz and R\"{o}ssler system as our test systems. The reservoir is trained to infer the last two variables from the first variable in input as provided in Equations (\ref{eq:lorenz} \& \ref{eq:rossler}). The reservoir multifunctions as two observers for two different chaotic systems.

\begin{figure}
\includegraphics[width = \linewidth]{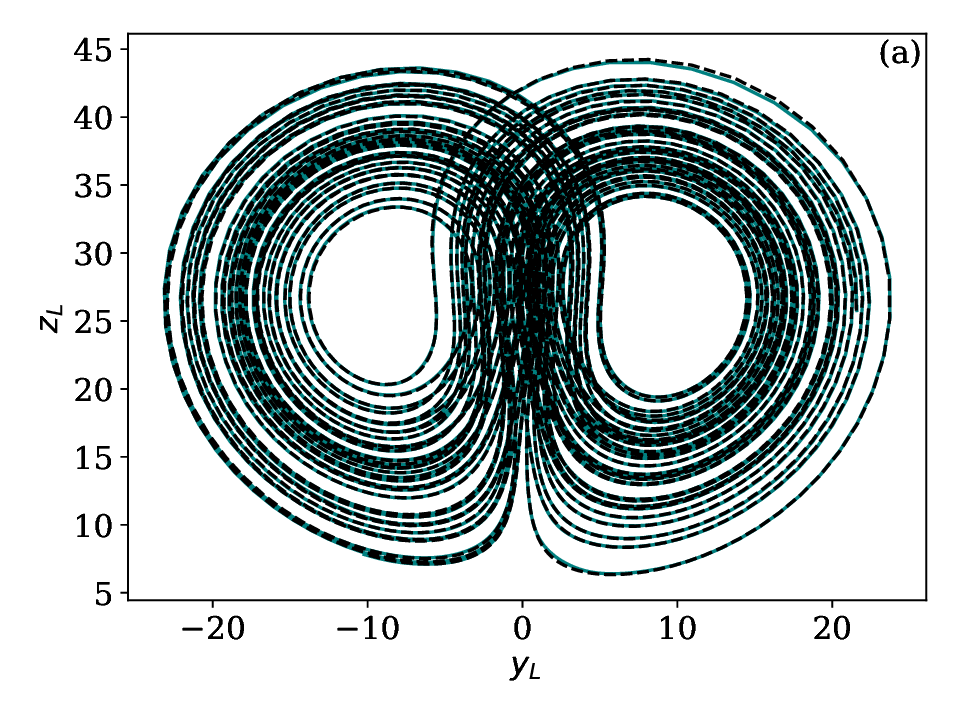}
\includegraphics[width = \linewidth]{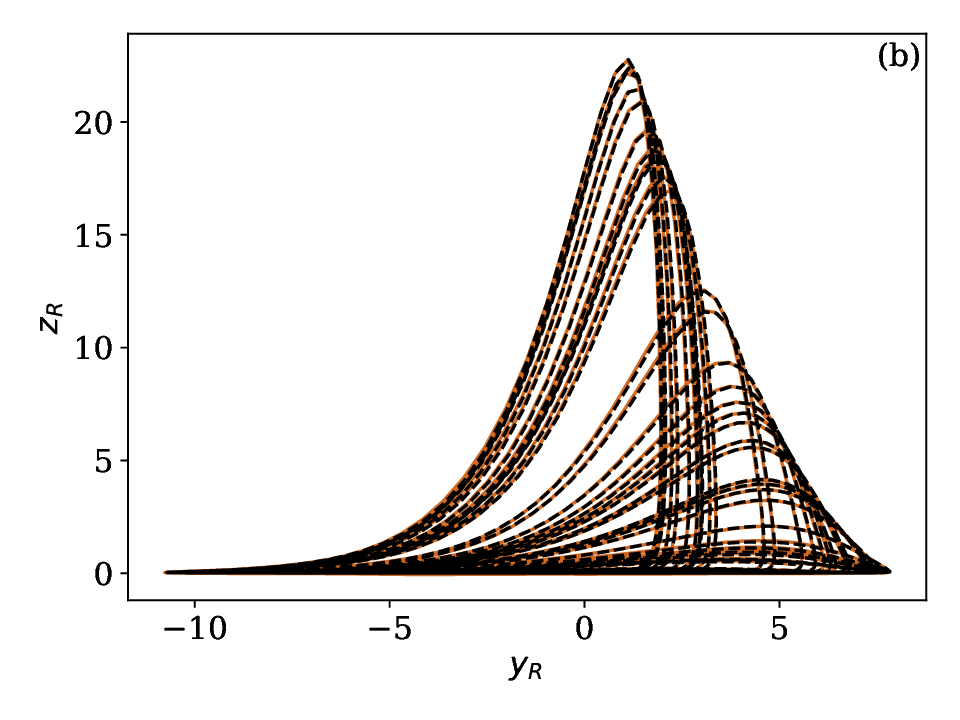}
\caption{\label{fig:obs} Plot of inferred variables in $y - z$ plane: (a) Lorenz attractor (b) R\"{o}ssler attractor. The dotted black lines represent the true trajectories of individual attractors; colored lines are the trajectories inferred by the multifunctional reservoir.}
\end{figure}

The performance is depicted in Figure \ref{fig:obs}, where two inferred variables are plotted in their phase plane. In each case, the inferred attractors are compared with their corresponding target trajectories. Task specific hyperparameters are set as $N^1 = N^2 = N_{res}/2$ and $\sigma^1 = \sigma^2 = 1.0$. Only the training data length are different as $T^1_{tr} = 10000$ and $T^2_{tr} = 20000$ for Lorenz and R\"{o}ssler systems, respectively. The training data was generated using Equations (\ref{eq:lorenz} \& \ref{eq:rossler}) with respective sampling rates $\Delta t = 0.01$ and $\Delta t = 0.05$.

\begin{table*}
\begin{tabular}{|l|c|c|c|c|c|}
\hline
\multicolumn{1}{|c|}{Hyperparameters} & Case {\bf A} & Case {\bf B} & Case {\bf C} & Case {\bf D} & Case {\bf E} \\ 
\hline 
\hline
Reservoir Size ($N_{res}$) & $2000$ & $3000$ & $3000$ & $3000$ & $500$  \\ 
\hline 
Reservoir Spectral Radius ($\rho$) & $0.5$ & $0.5$ & $0.5$ & $0.3$ & $0.9$  \\ 
\hline 
Reservoir Density ($d$)  & $0.2$ & $0.2$ & $0.2$ & $0.2$ & $0.01$ \\ 
\hline 
Leaking Parameter ($\alpha$) & $0.5$ & $0.5$ & $0.5$ & $0.5$ & $0.1$ \\ 
\hline 
Regularization Parameter ($\beta$) & $1\times 10^{-6}$ & $1\times 10^{-6}$ & $1\times 10^{-6}$ & $1\times 10^{-6}$ & $1\times 10^{-8}$ \\ 
\hline 
\end{tabular} 
\caption{Values of hyperparameters considered for different numerical experiments described in Sec. \ref{ssec:numsim1} - \ref{ssec:numsim5}, respectively.}
\label{tab:hype}
\end{table*}

\begin{figure}
\includegraphics[width = \linewidth]{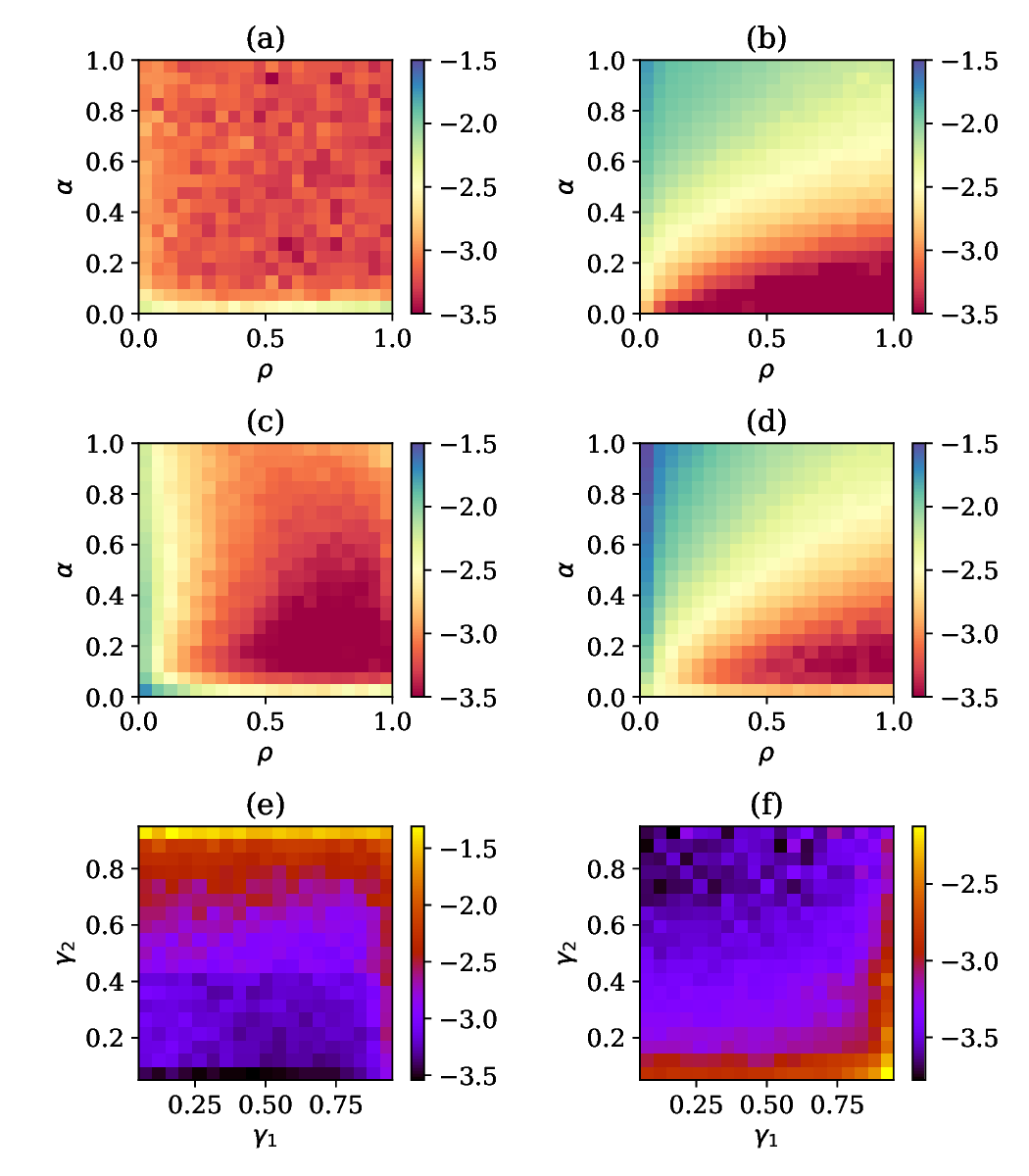}
\caption{\label{fig:comp} Plot of root-mean-squar errors (RMSE) for reservoir observer tasks for Lorenz (left column) and R\"{o}ssler (right column) systems, as described in Sec. \ref{ssec:numsim5}. Color axes represent  RMSE values in power of $10$. The first row (a \& b) depicts the variation in performance of mono-functional reservoirs in $\rho-\alpha$ space. Whereas, the second row (c \& d) represents that for the multifunctional reservoir trained with Lorenz and R\"{o}ssler data sets, keeping all other hyperparameters the same as the first row. The final row (e \& f) depicts the performance of individual tasks of the multifunctional reservoir in $\gamma_1-\gamma_2$ space.}
\end{figure}

\section{Discussions and Conclusion}\label{sec:disc}

 A compromise in performance is noticed when a mono-functional reservoir is trained for multifunctionality as shown in Figure \ref{fig:comp}. In the figure, multifunctional reservoir's performance (the second row) is compared with the mono-functional reservoirs performing individual tasks (the first row). For both cases, the size of low-error regions have shrunken, indicating a slight sacrifice in performance to achieve multifunctionality by a single reservoir. The effect of task specific hyperparameters on individual tasks is also analyzed. Figures \ref{fig:comp}(e-f) capture the performance reservoir observer for Lorenz and R\"{o}ssler systems respectively in $\gamma_1-\gamma_2$ space, where $N^1 = \gamma_1N_{res};~ N^2 = (1-\gamma_1)N_{res}$ and $T^1_{tr} = \gamma_2T_{tr};~ T^2_{tr} = (1-\gamma_2)T_{tr}$.

\begin{figure}
\includegraphics[width = \linewidth]{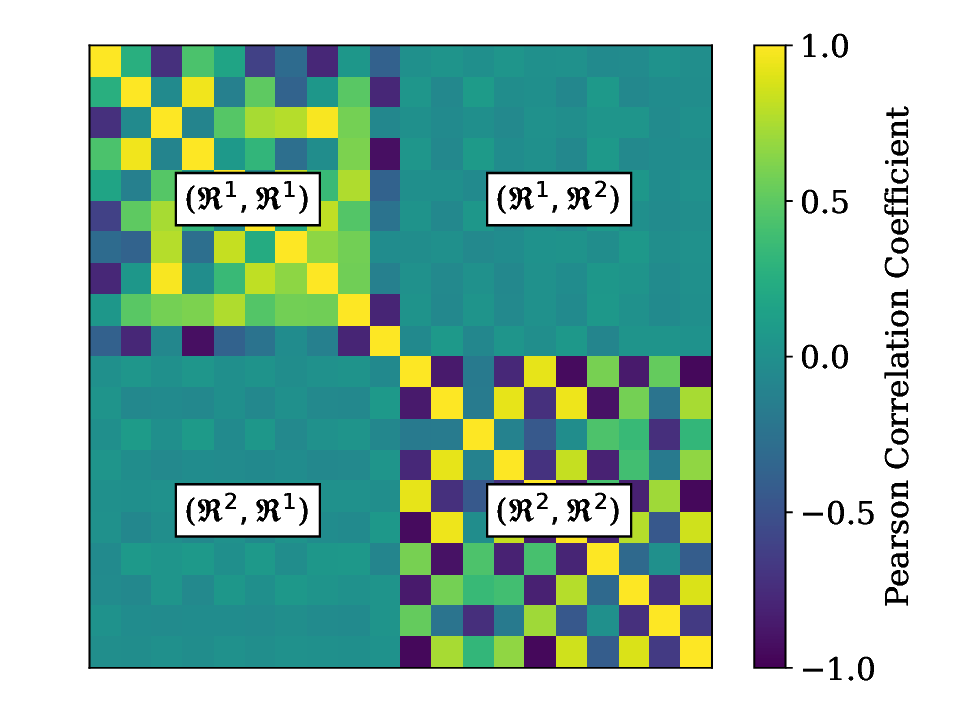}
\caption{\label{fig:corr} Pearson correlation among $10$ randomly selected reservoir nodes while trained to reconstruct Lorenz and R\"{o}ssler attractors. The ($\mathfrak{R}^1,\mathfrak{R}^1$) quadrant represents the correlation among the nodes while performing the Lorenz task. ($\mathfrak{R}^2,\mathfrak{R}^2$) represents that for R\"{o}ssler task. The off-diagonal quadrants represent the correlation of nodes while performing two different tasks.}
\end{figure}

The reservoir dynamics is analyzed by considering a few randomly selected nodes while performing different tasks in the proposed architecture. A Pearson correlation matrix is plotted in Figure \ref{fig:corr} for the case of reconstructing Lorenz and R\"{o}ssler attractors multifunctionally. The dynamics of each node is highly correlated (or anti-correlated) with other nodes while performing either of the tasks. However, the dynamics of the nodes for two different tasks are almost uncorrelated. This implies that the reservoir dynamics evolves into two completely uncorrelated attractors for two functions. For this case of two attractor climate reconstruction tasks, the reservoir can predict two climates starting from identical initial conditions. Thus, the whole setup during the closed-loop prediction phase works as a digital twin of a dynamical system that has both Lorenz and R\"{o}ssler attractors, accessible by configuring two input channels. Whereas, in the previous approach of multifunctionality in RC, the reservoir leads to different attractors depending upon initial conditions only. Configuring such a multistable reservoir is cumbersome, often leading to performance instability. On the other hand, in the proposed approach, different attractors are accessed by system parameter configuration (input connection matrices in this case). This provides greater flexibility in setting up a multifunctional RC. Moreover, the unparalleled performance of the approach is evident in the last section.

The study achieves a significant breakthrough in reservoir computing by developing an enhanced framework that enables multifunctionality beyond previous limitations. The proposed approach allows RC to predict multiple chaotic attractors, even when they share a common phase space—a challenge that prior methods struggled to address. Instead of relying on multistability within reservoir dynamics, the model introduces separate input channels for different tasks, making it more flexible and biologically relevant. This advancement enhances computational efficiency by allowing a single network to perform multiple functions without structural modifications, reducing training complexity and improving adaptability. 

However, the proposed method is able to handle tasks that are functionally similar in nature. For example, it is still not possible for a single reservoir to perform climate reconstruction as well as reservoir observer tasks. A possible reason is the requirement of significantly different reservoir properties for two tasks. Further research can be aimed towards making the scheme multimodal, handling tasks of different kinds by single reservoir. Nonetheless, the study showcases the effectiveness of this enhanced RC framework across various challenging dynamical scenarios, including climate reconstruction for Lorenz and R\"{o}ssler systems, handling attractors of different dimensions, reconstructing hyperchaotic systems, and serving as a reservoir observer. By expanding the scope of multifunctional RC, the study strengthens its applicability in robotics (for smooth task transitions), neuroscience (for modeling cognitive flexibility and motor control), and signal processing (for improved speech recognition and biomedical diagnostics). These achievements make reservoir computing more powerful, versatile, and capable of tackling complex nonlinear problems.

\section*{Conflict of Interest}
The authors have no conflict to disclose.

\section*{Data Availability}
The data that support the findings of this study are available from the corresponding author upon reasonable request.

\section*{Acknowledgment}
SM and KA acknowledge the financial support by JST Moonshot R\&D Grant Number JP-MJMS2021. KA also acknowledges Institute of AI and Beyond of UTokyo, and Cross-ministerial Strategic Innovation Promotion Program (SIP), the 3rd period of SIP , Grant Numbers JPJ012207, JPJ012425.

\section*{references}
\bibliography{citations}

\end{document}